\documentclass[11pt]{article}
\usepackage{latexsym,epic,eepic}
 \usepackage{graphicx}
\begin{document}
\newcommand{\2}{\vspace{0.2 cm}}
\newcommand{\dist}{{\rm dist}}
\newcommand{\diam}{{\rm diam}}
\newcommand{\rad}{{\rm rad}}
\newcommand{\dom}{\mbox{$\rightarrow$}}
\newcommand{\ldom}{\mbox{$\leftarrow$}}
\newcommand{\edom}{\mbox{$\leftrightarrow$}}
\newcommand{\qed}{\hfill$\diamond$}
\newcommand{\pf}{{\bf Proof: }}
\newtheorem{theorem}{Theorem}[section]
\newcommand{\ra}{\rangle}
\newcommand{\la}{\langle}
\newtheorem{lemma}[theorem]{Lemma}
\newtheorem{claim}[theorem]{Claim}
\newtheorem{definition}[theorem]{Definition}
\newtheorem{corollary}[theorem]{Corollary}
\newtheorem{observation}[theorem]{Observation}
\newtheorem{proposition}[theorem]{Proposition}
\newtheorem{conjecture}[theorem]{Conjecture}
\newtheorem{problem}[theorem]{Problem}
\newtheorem{question}[theorem]{Question}
\newtheorem{remark}[theorem]{Remark}
\newcommand{\beq}{\begin{equation}}
\newcommand{\eeq}{\end{equation}}
\newcommand{\UN}{{\rm UN}}
\newcommand{\MiP}{MinHOM($H$) }
\newcommand{\MaP}{MaxHOM($H$) }
\newcommand{\vecc}[1]{\stackrel{\leftrightarrow}{#1}}
\newcommand{\sdom}{\mbox{$\Rightarrow$}}
\newcommand{\nsdom}{\mbox{$\not\Rightarrow$}}
\newcommand{\eps}[2][1]{%
  \scalebox{#1}{\includegraphics{fig/hex.#2}}}
\newcommand{\labcontent}[2]{%
  \vbox{\halign{\hfil##\hfil\cr#1\cr(#2)\cr}}}
\newcommand{\labeps}[3][1]{\labcontent{\eps[#1]{#2}}{#3}}
\newcommand{\fline}[1]{\hbox to \hsize{\hfil#1\hfil}}

\title{Minimum Cost Homomorphisms to Reflexive Digraphs}

\date{}
\author{Arvind Gupta \ \ Pavol Hell \ \  Mehdi Karimi \ \ Arash Rafiey \\ \\
School of Computing Science \\ Simon Fraser University\\ Burnaby, B.C., Canada, V5A 1S6
\thanks{arvind@mitacs.ca , pavol@cs.sfu.ca , mmkarimi@cs.sfu.ca , arashr@cs.sfu.ca}
}
\maketitle

\begin{abstract}
For digraphs $G$ and $H$, a homomorphism of $G$ to $H$ is a mapping
$f:\ V(G)\dom V(H)$ such that $uv\in A(G)$ implies $f(u)f(v)\in
A(H)$. If moreover each vertex $u \in V(G)$ is associated with
costs $c_i(u), i \in V(H)$, then the cost of a homomorphism $f$ is
$\sum_{u\in V(G)}c_{f(u)}(u)$. For each fixed digraph $H$, the {\em
minimum cost homomorphism problem} for $H$, denoted MinHOM($H$), is
the following problem. Given an input digraph $G$, together with
costs $c_i(u)$, $u\in V(G)$, $i\in V(H)$, and an integer $k$, decide
if $G$ admits a homomorphism to $H$ of cost not exceeding $k$. Minimum
cost homomorphism problems encompass (or are related to) many well
studied optimization problems such as chromatic partition optimization
and applied problems in repair analysis. For undirected graphs the
complexity of the problem, as a function of the parameter $H$, is well
understood; for digraphs, the situation appears to be more complex, and
only partial results are known. We focus on the minimum cost homomorphism
problem for {\em reflexive} digraphs $H$ (every vertex of $H$ has a loop).
It is known that the problem MinHOM($H$) is polynomial time solvable if
the digraph $H$ has a {\em Min-Max ordering}, i.e., if its vertices can be
linearly ordered by $<$ so that $i<j, s<r$ and $ir, js \in A(H)$ imply that
$is \in A(H)$ and $jr \in A(H)$. We give a forbidden induced subgraph
characterization of reflexive digraphs with a Min-Max ordering; our
characterization implies a polynomial time test for the existence of
a Min-Max ordering. Using this characterization, we show that for a
reflexive digraph $H$ which does not admit a Min-Max ordering, the
minimum cost homomorphism problem is NP-complete, as conjectured
by Gutin and Kim. Thus we obtain a full dichotomy classification of the
complexity of minimum cost homomorphism problems for reflexive digraphs.
\end{abstract}

\section{Introduction and Terminology}\label{introsec}

For digraphs $G$ and $H$, a mapping $f:\ V(G)\dom V(H)$ is a
{\em homomorphism of $G$ to $H$} if $uv$ is an arc of $G$ implies
$f(u)f(v)$ is an arc of $H$. Let $H$ be a fixed digraph: the
{\em homomorphism problem} for $H$, denoted HOM$(H)$, asks whether
or not an input digraph $G$ admits a homomorphism to $H$. The {\em
list homomorphism problem} for $H$, denoted ListHOM$(H)$, asks whether
or not an input digraph $G$, with lists $L_u \subseteq V(H), u \in V(G)$,
admits a homomorphism $f$ to $H$ in which all $f(u) \in L_u, u \in V(G)$.

Suppose $G$ and $H$ are digraphs, and $c_i(u)$, $u\in V(G)$, $i\in V(H)$,
are real {\em costs}. The {\em cost of a homomorphism} $f$ of $G$ to $H$
is $\sum_{u\in V(G)}c_{f(u)}(u)$. If $H$ is fixed, the {\em minimum cost
homomorphism problem} for $H$, denoted MinHOM($H$), is the following problem.
Given an input digraph $G$, together with costs $c_i(u)$, $u\in V(G)$,
$i\in V(H)$, and an integer $k$, decide if $G$ admits a homomorphism to
$H$ of cost not exceeding $k$.

If the graph $H$ is {\em symmetric} (each $uv \in A(H)$ implies
$vu \in A(H)$), we may view $H$ as an undirected graph. In this way, we may
view the problem MinHOM($H$) as applying also to undirected graphs.

The minimum cost homomorphism problem was introduced, in the context
of undirected graphs, in \cite{gutinDAMlora}. There, it was motivated by a
real-world problem in defense logistics; in general, the problem seems to
offer a natural and practical way to model many optimization problems.
Special cases include for instance the list homomorphism problem
\cite{hell2003,hell2004} and the optimum cost chromatic partition problem
\cite{halld2001,jansenJA34,jiangGT32} (which itself has a number of
well-studied special cases and applications \cite{kroon1997,supowitCAD6}).

Our interest is in proving dichotomies: given a class of problems such as
HOM($H$), we would like to prove that for each digraph $H$ the problem
is polynomial-time solvable, or NP-complete. This is, for instance, the case
for HOM($H$) with undirected graphs $H$ \cite{hellJCT48}; in that case
it is known that HOM($H$) is polynomial time solvable when $H$ is
bipartite or has a loop, and NP-complete otherwise \cite{hellJCT48}.
This is a dichotomy {\em classification}, since we specifically classify
the complexity of the problems HOM($H$), depending on $H$.

For undirected graphs $H$, a dichotomy classification for the problem
MinHOM($H$) has been provided in \cite{mincostungraph}. (For ListHOM$(H)$,
consult \cite{pavol}.) Thus, the minimum cost homomorphism problem for
graphs has been handled, and interest shifted to directed graphs. The
first studies \cite{gutinDAM,gutinDO,yeo} focused on irreflexive digraphs
(no vertex has a loop), where dichotomies has been obtained for digraphs
$H$ such that $U(H)$ is a complete or complete multipartite graph. More
recently, \cite{gregorykim2} promoted the study of digraphs with loops
allowed; and, in particular, of reflexive digraphs. Dichotomy has been
proved for reflexive digraphs $H$ such that $U(H)$ is a complete graph,
or a complete multipartite graph without digons \cite{gregorykim,gregorykim3}.
In this paper, we give a full dichotomy classification of the complexity of
MinHOM($H$) for reflexive digraphs; this is the first dichotomy result
for a general class of digraphs - our only restriction is that the digraphs
are reflexive. The dichotomy classification we prove verifies a conjecture
of Gutin and Kim \cite{gregorykim}. (Partial results on ListHOM$(H)$ for
digraphs can be found in  \cite{Hellpowerdigraph, feder, Hellpartitioning,
federmanuscript,Federcycle,helltree,Zhou}.

Let $H$ be any digraph. An arc $xy \in A(H)$ is {\em symmetric} if $yx \in A(H)$;
the digraph $H$ is {\em symmetric} if each arc of $H$ is symmetric. Otherwise,
we denote by $S(H)$ the {\em symmetric subgraph} of $H$, i.e., the undirected
graph with $V(S(H))=V(H)$ and $E(S(H))=\{uv : uv \in A(H)$ and $vu \in A(H)\}$.
We also denote by $U(H)$ the {\em underlying graph} of $H$, i.e., the undirected
graph with $V(U(H))=V(H)$ and $E(U(H))=\{uv : uv \in A(H)$ or $vu \in A(H)\}$.
If $H$ is a reflexive digraph, then both $S(H)$ and $U(H)$ are reflexive graphs.
Finally, we denote by $B(H)$ the bipartite graph obtained from $H$ as follows.
Each vertex $v$ of $H$ gives rise to two vertices of $B(H)$ - a {\em white} vertex
$v'$ and a {\em black}  vertex $v''$; each arc $vw$ of $H$ gives rise to an edge
$v'w''$ of $B(H)$. Note that if $H$ is a reflexive digraph, then all edges $v'v''$ are
present in $B(H)$. The {\em converse} of $G$ is the digraph obtained from $G$ by
reversing the directions of all arcs.

\begin{figure}
   \begin{center}
      \includegraphics [scale =.22, angle =270]{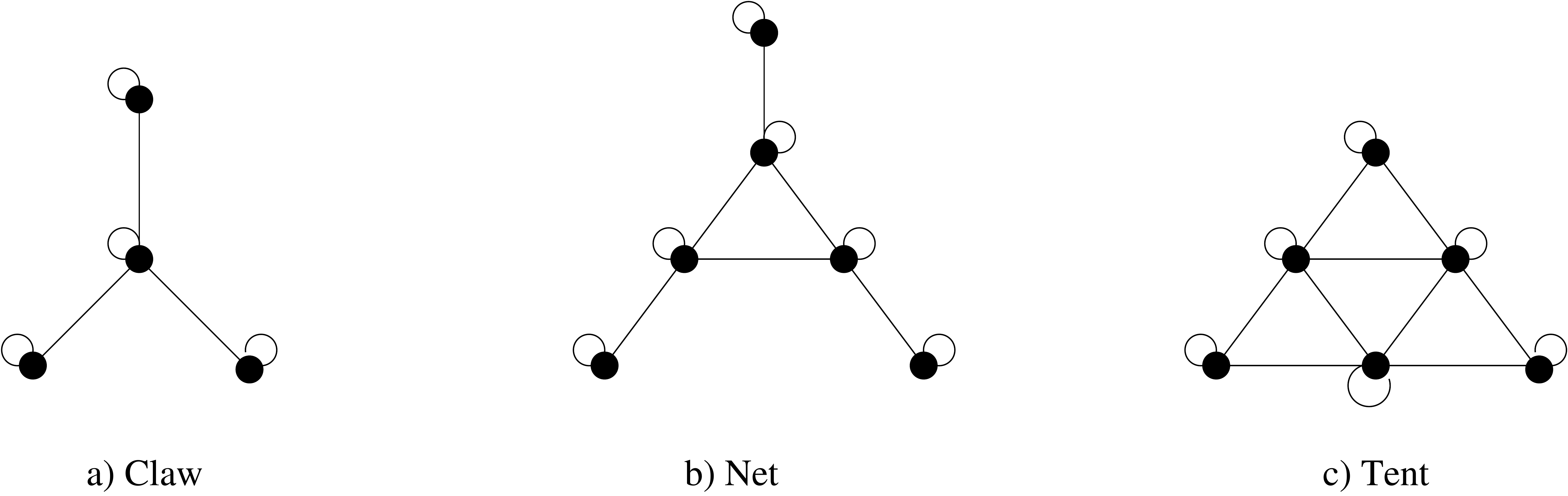}
   \end{center}
   \caption{The claw, the net, and the tent.}
   \label{nettentclaw}
 \end{figure}

We say that an undirected graph $H$ is a {\em proper interval graph}
if there is an inclusion-free family of intervals $I_v, v \in V(H),$
such that $vw \in E(H)$ if and only if $I_v$ intersects $I_w$. Note
that by this definition proper interval graphs are reflexive. Wegner
proved \cite{Wegner} that a reflexive graph $H$ is a proper interval
graph if and only if it does not contain an induced cycle $C_k$,
with $k \geq 4$, or an induced claw, net, or tent, as given in
Figure \ref{nettentclaw}.

\begin{figure}
   \begin{center}
      \includegraphics [scale =.18, angle =270]{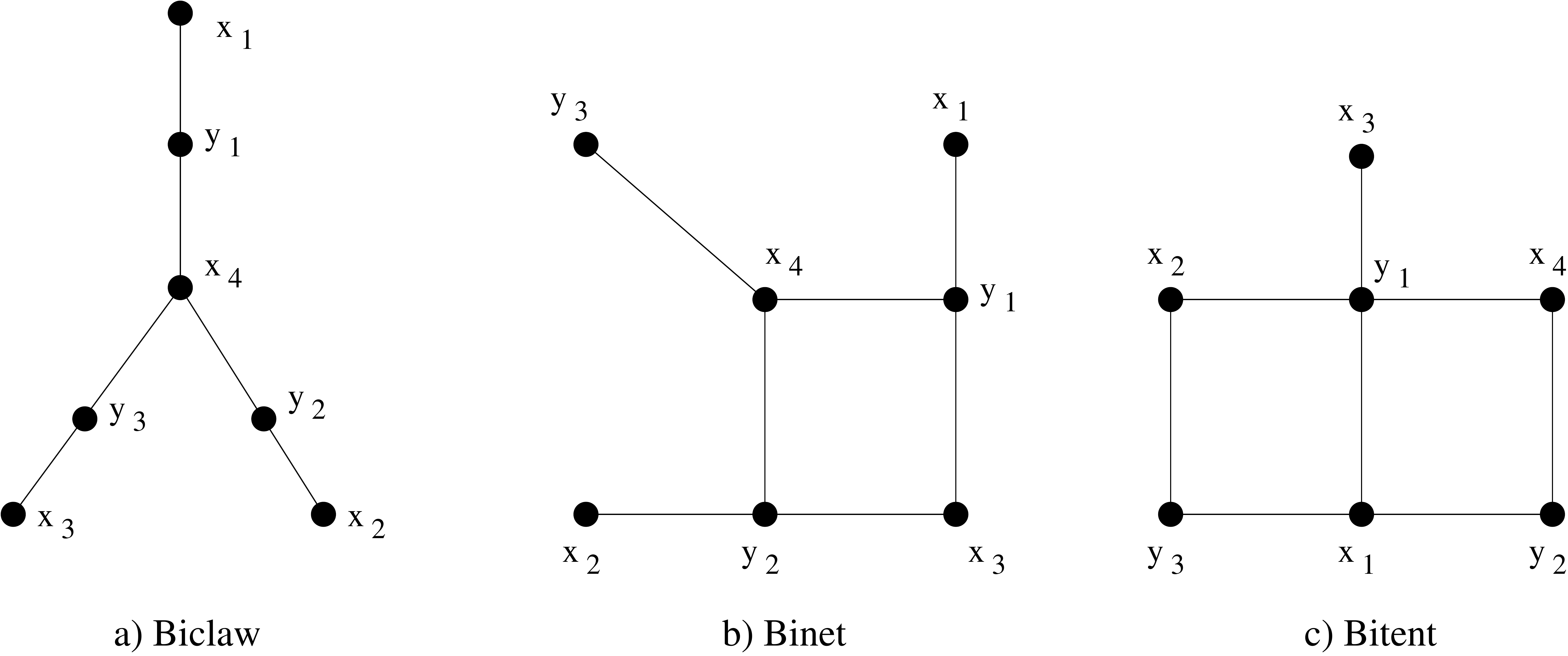}
   \end{center}
   \caption{The biclaw, the binet, and the bitent.}
   \label{bipartitenettentclaw}
 \end{figure}

 We say that a bipartite graph $H$ (with a fixed bipartition into white and black
 vertices) is a {\em proper interval bigraph} if there are two inclusion-free families
 of intervals $I_v$, for all white vertices $v$, and $J_w$ for all black vertices $w$,
 such that $vw \in E(H)$ if and only if $I_v$ intersects $J_w$. By this definition
proper interval bigraphs are irreflexive and bipartite. A Wegner-like characterization
(in terms of forbidden induced subgraphs) of proper interval bigraphs is given in
\cite{hellhuang2004}: $H$ is a proper interval bigraph if and only if it does not
contain an induced cycle  $C_{2k}$, with $k \geq 3$, or an induced biclaw, binet,
or bitent, as given in Figure \ref{bipartitenettentclaw}.

A linear ordering $<$ of $V(H)$ is a {\em Min-Max ordering} if $i<j, s<r$ and
$ir, js \in A(H)$ imply that $is \in A(H)$ and $jr \in A(H)$. For a reflexive digraph
$H$, it is easy to see that $<$ is a Min-Max ordering if and only if for any $j$
between $i$ and $k$, we have $ik \in A(H)$ imply $ij, jk \in A(H)$. For a bipartite
graph $H$ (with a fixed bipartition into white and black vertices), it is easy to see
that $<$ is a Min-Max ordering if and only if $<$ restricted to the white vertices,
and $<$ restricted to the black vertices satisfy the condition of Min-Max orderings,
i.e., $i<j$ for white vertices, and $s<r$ for black vertices, and $ir, js \in A(H)$,
imply that $is \in A(H)$ and $jr \in A(H)$). A {\em bipartite Min-Max ordering}
is an ordering $<$ specified just for white and for black vertices.

It is known that if $H$ admits a Min-Max ordering, then the problem
MinHOM($H$) is polynomial time solvable \cite{gutinDAM}, see also
\cite{cohenJAIR22,khana}; however, there are digraphs with polynomial
MinHOM($H$) which do not have Min-Max ordering \cite{gutinDO}.
For undirected graphs, all $H$ without a Min-Max ordering yield an
NP-complete MinHOM($H$) \cite{mincostungraph}; moreoever, having
a Min-Max ordering can be characterized by simple forbidden induced
subgraphs, and recognized in polynomial time \cite{mincostungraph}.
In particular, a {\em reflexive} graph admits a Min-Max ordering if
and only if it is a proper interval graph, and a {\em bipartite}
graph admits a Min-Max ordering if and only if it is a proper
interval bigraph \cite{mincostungraph}.

We shall give a combinatorial description of reflexive digraphs with Min-Max ordering,
in terms of forbidden induced subgraphs. Our  characterization yields a polynomial time
algorithm for the existence of a Min-Max ordering in a reflexive digraph. It also allows us
to complete a dichotomy classification of MinHOM($H$) for reflexive digraphs $H$, by
showing that all problems MinHOM($H$) where $H$ does not admit a Min-Max ordering
are NP-complete. This verifies a conjecture of Gutin and Kim in \cite{gregorykim}.

\section{Structure and Forbidden Subgraphs }\label{psec}

Since both reflexive and bipartite graphs admit a characterization of existence of
Min-Max orderings by forbidden induced subgraphs, our goal will be accomplished
by proving the following theorem. It also implies a polynomial time algorithm to test
if a reflexive digraph has a Min-Max ordering.

\begin{theorem}\label{minmax}
A reflexive digraph $H$ has a Min-Max ordering if and only if

\begin{itemize}
\item $S(H)$ is a proper interval graph, and

\item $B(H)$ is a proper interval bigraph, and

\item $H$ does not contain an induced subgraph isomorphic to $H_i$ with $i=1,2,3,4,5,6$.
\end{itemize}
\end{theorem}

\begin{figure}
   \begin{center}
     \includegraphics [scale =.22, angle =270]{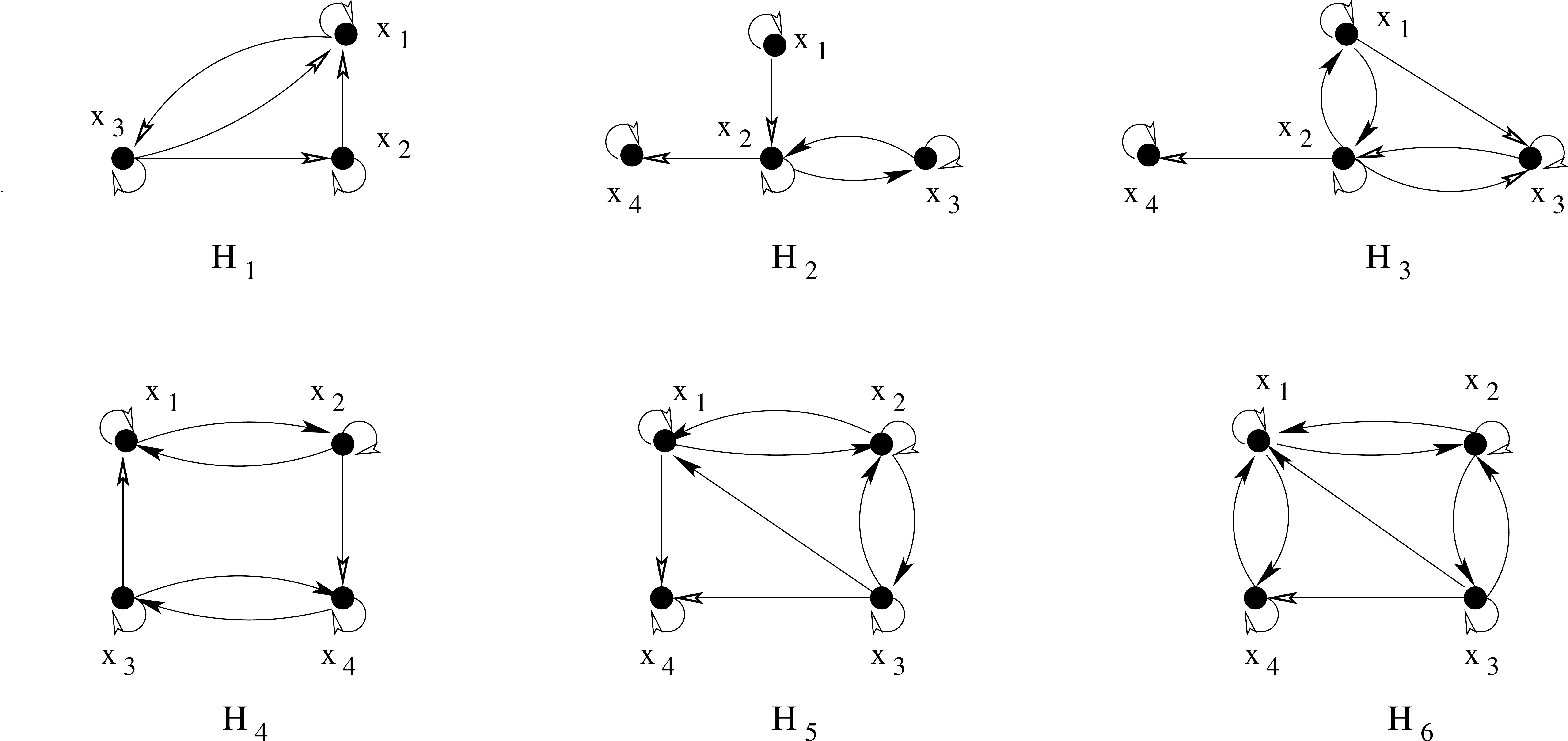}
   \end{center}
   \caption{The obstructions $H_i$ with $i=1,2,3,4,5,6$.}
   \label{O_r}
 \end{figure}

The digraphs $H_i$ are depicted in Figure \ref{O_r}. The resulting forbidden subgraph
characterization is summarized in the following corollary. Note that forbidden subgraphs
in $S(H)$ directly describe forbidden subgraphs in $H$, and it is easy to see that each
forbidden induced subgraph in $B(H)$ can also be translated to a small family of forbidden
induced subgraphs in $H$.

\begin{corollary}
A reflexive digraph $H$ has a Min-Max ordering if and only if $S(H)$ does not
contain an induced $C_k, k \geq 4$, or claw, net, or tent, $B(H)$ does not contain
an induced $C_{2k}, k \geq 3,$ or biclaw, binet, or bitent, and $H$ does not
contain an induced $H_i$ with $i=1,2,3,4,5,6$.
\end{corollary}

We proceed to prove the Theorem.

\pf Suppose first that $<$ is a Min-Max ordering $<$ of $H$. It is
easily seen that $<$ is also a Min-Max ordering of $S(H)$, and that
$<$ applied separately to the corresponding white and black vertices
of $B(H)$ is a bipartite Min-Max ordering of $B(H)$. To complete the
proof of necessity, we now claim that none of the digraphs $H_i,
i=1,2,3,4,5,6$ admits a Min-Max ordering. We only show this for
$H_3$, the proofs of the other cases being similar. Suppose that $<$
is a Min-Max ordering of $H_3$. For the triple $x_1, x_2, x_3$, we
note that $x_2$ must be between $x_1$ and $x_3$ in the ordering $<$,
as otherwise we would have $x_1x_3 \in E(S(H))$ . Without loss of
generality assume that $x_1<x_2<x_3$. Since $x_1$ and $x_4$ are
independent and $x_1x_2 \in E(S(H))$, we must have $x_4>x_1$. A
similar argument yields $x_4<x_3$; however, $x_1 < x_4 < x_3$ is
impossible, as $x_1x_3 \in A(H)$ but $x_1x_4 \not\in A(H)$.

To prove the sufficiency of the three conditions, we shall prove the
following claim.

\begin{lemma}
If $S(H)$ has a Min-Max ordering and $B(H)$ has a bipartite Min-Max ordering,
then either $H$ has a Min-Max ordering, or $H$ contains an induced $H_i$
(or its converse) for some $i=1,2,3,4,5,6$.
\end{lemma}

\pf Suppose $<$ is a bipartite Min-Max ordering of $B(H)$. A pair $u, v$ of
vertices of $H$ is {\em proper for $<$} if $u' < v'$ if and only if $u'' < v''$ in
$B(H)$. We say a bipartite Min-Max ordering $<$ is {\em proper} if all pairs
$u, v$ of $H$ are proper for $<$. If $<$ is a proper bipartite Min-Max ordering,
then we can define a corresponding ordering $\prec$ on the vertices of
$H$, where $u \prec v$ if and only if $u' < v'$ (which happens if and only if
$u'' < v''$). It is easy to check that $\prec$ is now a Min-Max ordering of $H$.

Suppose, on the other hand, that the bipartite Min-Max ordering $<$
on $B(H)$ is not proper. Thus there are vertices $v', u'$ such that
$v' < u'$ and $u'' < v''$. Suppose there is no vertex $s'$ such that
$s'v'' \in E(B(H))$, $s'u'' \not\in E(B(H))$: then we can
exchange the position of $v''$ and $u''$ in $<$ and still have a
bipartite Min-Max ordering. Furthermore, this exchange strictly
increases the number of proper pairs in $H$: any $w$ with $u''
< w'' < v''$ and $u' < w'$ creates a new improper pair $u, w$
but also creates a new proper pair $v, w$ (and the pair $u, v$
is also a new proper pair). Analogously, if there is no vertex $t''$
such that $u't'' \in E(B(H))$, $v't'' \not\in E(B(H))$, we can
exchange $u',v'$ and increase the number of proper pairs in $H$.
Suppose we have performed all exchanges until we reached a
bipartite Min-Max ordering $<$ which admits no more exchanges.
Then there are two possibilities: either $<$ is now proper, and $H$
admits a Min-Max ordering as above, or $<$ is still not proper, and
one of the following two cases must occur (up to symmetry):

{\bf Case 1: } $s'v'', v't'' \in E(B(H))$ and $s'u'', u't'' \not\in E(B(H))$.

It is easy to see that since $<$ is a bipartite Min-Max ordering,
we must have $u' < s' $ and $t'' < u''$. (Note that means that
$s'' \neq t''$.) Since $u'u'', v'v'' \in E(B(H))$, by the same argument
we must have $u'v'', v'u'' \in E(B(H))$; and similarly we obtain
$s't''\not\in E(B(H))$. If both $v's''$ and $t'v''$ are edges of $B(H)$
then $u, v, s, t$ induce a claw in $S(H)$: indeed in $B(H)$, we
have the edges $v't'', t'v'', v'u'', u'v'', v's'', s'v''$ and the non-edges
$u't'', s'u'', s't''$. This is a contradiction, as $S(H)$ is assumed to
have a Min-Max ordering, i.e., be a proper interval graph.

If neither $v's''$ nor $t'v''$ is an edge of $B(H)$, then if $u's''$ is an
edge of $B(H)$, then $s, v, u$ induce a copy of $H_1$ in $H$, and if $,t'u''$
is an edge of $B(H)$, then $t, v, u$ induce a copy of $H_1$. Thus consider the
case when $u's'', t'u'' \not\in E(B(H))$. If $t's'' \in E(B(H))$, then
$s' ,s'', t', t'', v', v''$ would induce a copy of $C_{6}$ in $B(H)$, contrary
to our assumption that $B(H)$ has a bipartite Min-Max ordering, i.e., is a
proper interval bigraph. Thus $t's'' \not\in E(B(H))$ and $t, s, v, u$
induce a copy of $H_2$ in $H$.

If only one of $v's''$ or $t'v''$ is an edge of $B(H)$, assume first that
$v's''\in E(B(H))$ and $t'v''\not\in E(B(H))$. If $t'u''$ is an edge of $B(H)$,
then $t, v, u$ induce a copy of $H_1$ in $H$, and if $t's''$ is an edge of
$B(H)$, then $t, v, s$ similarly induce a copy of $H_1$; thus asume that
$t'u'',t's''\not\in E(B(H))$. Note that $u's''\in E(B(H))$, else the vertices
$u', u'', v', t'', t', s'', s'$ would induce a biclaw in $B(H)$, contrary to
$B(H)$ being a proper interval bigraph. It now follows that $s, t, u, v$
induce a copy of $H_3$ in $H$. If $v's''\not\in E(B(H))$ and $t'v''\in E(B(H))$,
the proof is similar, except we obtain copies of $H_1$ and the converse
of $H_3$.

{\bf Case 2: }$s'v'',u't'' \in E(B(H))$ and $s'u'',v't'' \not\in E(B(H))$.

We again easily observe that we must have $u' <s'', v''< t''$, and
$u'v'',v'u'' \in E(B(H))$. If $s''=t''$ we obtain a copy of $H_1$
induced by $u, v, s$ in $H$; hence we assume that $s'' \neq t''$.
Suppose first that $u's'',t'v'' \not\in E(B(H))$. We have $s' < t'$
and $t'' < s''$, and so $t's'',s't'' \in A(H)$, implying that
$u, v, s, t$ induce a copy of $H_4$ in $H$. Suppose next that
both $t'v'',u's'' \in E(B(H))$. If $v's''$ is not an edge of $B(H)$,
vertices $u, v, s$ induce a copy of $H_1$ in $H$, and if $t'u''$
is not an edge of $B(H)$, vertices $u, v, t$ induce a copy of $H_1$
in $H$. Thus we have $v's'',t'u'' \in E(B(H))$. Now we have $t' < s'$
and $s'' < t''$, and hence $t's'',s't'' \in E(B(H))$. This is impossible,
since $u, v, s, t$ would induce a copy of $C_4$ in $S(H)$.
Finally,, if only one of $t'v'', u's''$ as an edge of $B(H)$, say
$u's'' \in E(B(H))$ and $t'v'' \not\in E(B(H))$ (the other case is
symmetric), then with the same argument as above, $v's''\in E(B(H))$,
$s't'' \in E(B(H))$, and $s, t, u, v$ induce (depending on which of
the pairs $t'u'', t's''$ are edges of $B(H)$) one of $H_1, H_5$ (or its
converse), or $H_6$ (or its converse).
\qed

\section{Complexity}

If $H$ has a Min-Max ordering, then MinHOM($H$) is polynomial time
solvable \cite{gutinDAM} see also \cite{cohenJAIR22,khana}. Now using
our forbidden induced subgraph characterization we can prove that
reflexive digraphs $H$ without a Min-Max ordering yield NP-complete
MinHOM($H$) problems. Note that we already know that MinHOM($S(H)$)
is NP-complete if $S(H)$ is not a proper interval graph, and
MinHOM($B(H)$) is NP-complete if $B(H)$ is not a proper interval
bigraph \cite{gutinDAM}. We begin with a few simple observations.
They first one is easily proved by setting up a natural polynomial time
reduction from MinHOM($B(H)$) to $MinHOM(H)$ \cite{gregorykim2}.

\begin{proposition}\cite{gregorykim2} \label{reduction1}
If MinHOM($B(H)$) is NP-complete, then $MinHOM(H)$ is also NP-complete.
\qed
\end{proposition}

The next two observations are folklore, and proved by obvious reductions,
cf. \cite{gregorykim}.

\begin{proposition} \label{symm}
If MinHOM($S(H)$) is NP-complete, then MinHOM($H$) is also NP-complete.
\qed
\end{proposition}

\begin{proposition}\label{subdigraph}
Let $H'$ be an induced subgraph of the digraph $H$. If MinHOM($H'$)
is NP-complete then MinHOM($H$) is NP-complete.
\qed
\end{proposition}

We now continue to prove that MinHOM($H$) is NP-complete for digraphs
$H=H_1,\ldots,H_6$. Let $\mathcal I$ denote the following decision
problem: given a graph $X$ and an integer $k$, decide whether or
not $X$ contains an independent set of $k$ vertices. This problem
has been useful for proving NP-completeness of minimum cost
homomorphism problems for undirected graphs \cite{mincostungraph},
and we use it again for digraphs.

\begin{proposition} \cite{mincostungraph} \label{MaxIndepNPhard}
The problem $\mathcal I$ is NP-complete, even when restricted to
three-colourable graphs (with a given three-colouring).
\qed
\end{proposition}

We denote by ${\mathcal I}_3$ the restriction of $\mathcal I$ to graphs
with a given three-colouring. In the following Lemmas, we give a
polynomial time reductions from ${\mathcal I}_3$. Note that all problems
MinHOM($H$) are in NP. The NP-completeness of MinHOM($H_1$)
follows from \cite{gregorykim}, Lemma 2-4.

\begin{lemma}\label{h2}
The problem $MinHOM(H_2)$ is NP-complete.
\end{lemma}

\pf We now construct a polynomial time reduction from ${\mathcal
I}_3$ to MinHOM($H_2$). Let $X$ be a graph whose vertices are
partitioned into independent sets $U, V, W$, and let $k$ be a given
integer. We construct an instance of MinHOM($H_2$) as follows: the
digraph $G$ is obtained from $X$ by replacing each edge $uv$ of $X$
with $u \in U, v \in V$ by an arc $uv$, replacing each edge $uw$ of
$X$ with $u \in U, w \in W$ by an arc $uw$, and replacing each edge
$vw$ of $X$ with $v \in V, w \in W$ by an arc $wv$. The costs are
defined by (writing for simplicity $c_i(y)$ for $c_{x_i}(y)$)
$c_1(u)=0, c_2(u)=1$ for $u \in U$, $c_4(v)=0, c_2(v)=1$ for $v \in
V$, and $c_3(w)=0, c_2(w)=1$, for $w \in W$. All other
$c_i(y)=|V(X)|$.

We now claim that $X$ has an independent set of size $k$ if and
only if $G$ admits a homomorphism to $H_2$ of cost $|V(X)|-k$.
Let $I$ be an independent set in $G$. We can  define a mapping
$f : V(G) \rightarrow V(H_2)$ as follows:

\begin{itemize}
\item
$f(u)=x_1$ for $u \in U \cap I$ and $f(u)=x_2$ for $u \in U - I$
\item
$f(v)=x_4$ for $v \in V \cap I$ and $f(v)=x_2$ for $v \in V - I$
\item
$f(w)=x_3$ for $w \in W \cap I$ and $f(w)=x_2$ for $w \in W - I$
\end{itemize}

This is a homomorphism of $G$ to $H_2$ of cost $|V(X)|-k$.

Let $f$ be a homomorphism of $G$ to $H_2$ of cost $|V(X)|-k$.
If $k\leq 0$ then we are trivially done so assume that $k>0$,
which implies that all individual costs are either zero or one.
Let $I=\{y \in V(X) \mbox{ $|$ } c_{f(y)}(y)=0 \}$ and note that
$|I| \geq k$. It can be seen that $I$ is an independent set in $G$,
as if $uv \in E(G)$, where $u \in I \cap U$ and $v \in I \cap V$
then $f(u)=x_1$ and $f(v)=x_4$, contrary to $f$ being a homomorphism.
\qed

\begin{lemma}\label{h3}
$MinHOM(H_3)$ is NP-complete.
\end{lemma}
\pf The reduction from the proof of Lemma \ref{h2} also applies here. \qed

\begin{lemma} \label{h4}
$MinHOM(H_4)$ is NP-complete.
\end{lemma}

\pf We now construct a polynomial time reduction from ${\mathcal I}_3$
to MinHOM($H_4$). Let $X$ be a graph whose vertices are partitioned
into independent sets $U, V, W$, and let $k$ be a given integer. An instance
of MinHOM($H_4$) is formed as follows: the digraph $G$ is obtained from
$X$ by replacing each edge $uv$ of $X$ with $u \in U, v \in V$ by an arc $vu$,
replacing each edge $uw$ of $X$ with $u \in U, w \in W$ by a directed path
$um_{uw}w$, and replacing each edge $vw$ of $X$ with $v \in V, w \in
W$ by a directed path $vm_{vw}w$. The costs are defined by $c_1(u)=1,
c_3(u)=0$ for $u \in U$; $c_2(v)=0, c_3(v)=1$ for $v \in V$; $c_4(w)=0,
c_1(w)=1$ for $w \in W$; $c_3(m_{uw})=c_4(m_{uw})=|V(X)|$ for each
edge $uw$ of $X$ with $u \in U, w \in W$; $c_2(m_{vw})=c_4(m_{vw})=|V(X)|$
for each edge $vw$ of $X$ with $v \in V, w \in W$; and $c_i(m)=0$ for any
other vertex $m \in V(G)-V(X)$, and $c_i(y)=|V(X)|$ for any other vertex
$y \in V(X)$.

We now claim that $X$ has an independent set of size $k$ if and only
if $G$ admits a homomorphism to $H_4$ of cost $|V(X)|-k$. Let $I$ be
an independent set in $G$. We can  define a mapping $f : V(G)
\rightarrow V(H_2)$ as follows:

\begin{itemize}
\item
$f(u)=x_3$ for $u \in U \cap I$ and $f(u)=x_1$ for $u \in U - I$
\item
$f(v)=x_2$ for $v \in V \cap I$ and $f(v)=x_3$ for $v \in V - I$
\item
$f(w)=x_4$ for $w \in W \cap I$ and $f(w)=x_1$ for $w \in W - I$
\item
$f(m_{uw})=x_2$ when $f(u)=x_1$, and $f(m_{uw})=x_1$ when $f(u)=x_3$
for each edge $uw$ of $X$ with $u \in U, w \in W$
\item
$f(m_{vw})=x_3$ when $f(w)=x_4$ and $f(m_{vw})=x_1$ when $f(w)=x_1$
for each edge $vw$ of $X$ with $v \in V, w \in W$
\end{itemize}

This is a homomorphism of $G$ to $H_4$ of cost $|V(X)|-k$.

Let $f$ be a homomorphism of $G$ to $H_4$ of cost $|V(X)|-k$. We may
again assume that all individual costs are either zero or one. Let $I=\{y
\in V(X) \mbox{ $|$ } c_{f(y)}(y)=0 \}$ and note that $|I| \geq k$.
It can be again seen that $I$ is an independent set in $G$, as if $uw \in
E(G)$, where $u \in I \cap U$ and $w \in I \cap V$ then $f(u)=x_3$
and $f(w)=x_4$, thus, $f(m_{uw})=x_3$ or $f(m_{uw})=x_4$. However,
the cost of homomorphism is greater than $|V(X)|$, a contradiction.
The other cases can also be treated similarly. \qed

\begin{lemma}\label{h5}
$MinHOM(H_5)$ is NP-complete.
\end{lemma}

\pf We similarly construct a polynomial time reduction from
${\mathcal I}_3$ to MinHOM($H_5$): this time the digraph $G$ is
obtained from $X$ by replacing each edge $uv$ of $X$ with $u \in U,
v \in V$ by an arc $uv$; replacing each edge $uw$ of $X$ with $u \in
U, w \in W$ by arcs $um_{uw}, wm_{uw}$; and replacing each edge $wv$
of $X$ with $w \in W, v \in V$ by a directed path $wm_{wv}v$. The
costs are $c_1(u)=1, c_2(u)=0$ for $u \in U$; $c_2(v)=1, c_4(v)=0$
for $v \in V$; $c_3(w)=1, c_1(w)=0$ for $w \in W$;
$c_1(m_{uw})=c_2(m_{uw})=|V(X)|$ for each edge $uw$ of $X$ with $u
\in U, w \in W$; $c_1(m_{wv})=c_4(m_{wv})=|V(X)|$ for each edge $wv$
of $X$ with $w \in W, v \in V$; $c_i(m)=0$ for any other vertex $m
\in V(G)-V(X)$, and $c_i(y)=|V(X)|$ for any other vertex $y \in
V(X)$.

We again claim that $X$ has an independent set of size $k$ if and only
if $G$ admits a homomorphism to $H_5$ of cost $|V(X)|-k$. Let $I$ be
an independent set in $G$. We can  define a mapping $f : V(G)
\rightarrow V(H_2)$ by $f(u)=x_2$ for $u \in U \cap I$ and $f(u)=x_1$
for $u \in U - I$; $f(v)=x_4$ for $v \in V \cap I$ and $f(v)=x_2$ for $v \in V - I$;
$f(w)=x_1$ for $w \in W \cap I$ and $f(w)=x_3$ for $w \in W - I$; $f(m_{uw})=x_3$
when $f(u)=x_2$, and $f(m_{uw})=x_4$ when $f(u)=x_1$, for each edge
$uw$ of $X$ with $u \in U, w \in W$; $f(m_{wv})=x_3$ when $f(w)=x_3$
and $f(m_{wv})=x_2$ when $f(w)=x_1$, for each edge $wv$ of $X$ with
$w \in W, v \in V$. This is a homomorphism of $G$ to $H_5$ of cost $|V(X)|-k$.

Let $f$ be a homomorphism of $G$ to $H_5$ of cost $|V(X)|-k$. Assuming
again that all individual costs are either zero or one, let $I=\{y
\in V(X) \mbox{ $|$ } c_{f(y)}(y)=0 \}$ and note that $|I| \geq k$.
It can be seen that $I$ is an independent set in $G$, as if $uw \in
E(G)$, where $u \in I \cap U$ and $w \in I \cap V$ then $f(u)=x_2$
and $f(w)=x_1$, thus, $f(m_{uw})=x_1$ or $f(m_{uw})=x_2$. However,
the cost of homomorphism is greater than $|V(X)|$, a contradiction.
The other cases can also be treated similarly. \qed

\begin{lemma}\label{h6}
$MinHOM(H_6)$ is NP-complete.
\end{lemma}

\pf The proof is again similar, letting the digraph $G$ be obtained from
$X$ by replacing each edge $uv$ of $X$ with $u \in U, v \in V$ by an
arc $uv$; replacing each edge $uw$ of $X$ with $u \in U, w \in W$ by
a directed path $um_{uw}w$; and replacing each edge $vw$ of $X$
with $v \in V, w \in W$ by an arc $wv$. The costs are defined by
$c_1(u)=0, c_2(u)=1$ for $u \in U$; $c_3(v)=0, c_1(v)=1$ for $v \in
V$; $c_4(w)=0, c_3(w)=1$; $c_1(m_{uw})=c_4(m_{uw})=|V(X)|$
for each edge $uw$ of $X$ with $u \in U, w \in W$; and letting
$c_i(m)=0$ for any other vertex $m \in V(G)-V(X)$, and $c_i(y)=|V(X)|$
for any other vertex $y \in V(X)$.

It can again be seen that $X$ has an independent set of size $k$ if and
only if $G$ admits a homomorphism to $H_6$ of cost $|V(X)|-k$: lettin
$I$ be an independent set in $G$, we  define a mapping $f : V(G)
\rightarrow V(H_2)$ by $f(u)=x_1$ for $u \in U \cap I$ and
$f(u)=x_2$ for $u \in U - I$; $f(v)=x_3$ for $v \in V \cap I$ and
$f(v)=x_1$ for $v \in V - I$; $f(w)=x_4$ for $w \in W \cap I$ and
$f(w)=x_3$ for $w \in W - I$; $f(m_{uw})=x_3$ when $f(u)=x_2$ and
$f(m_{uw})=x_2$ when $f(u)=x_1$ for each edge $uw$, $u \in U, w \in
W$. This is a homomorphism of $G$ to $H_6$ of cost $|V(X)|-k$.

Let $f$ be a homomorphism of $G$ to $H_6$ of cost $|V(X)|-k$ and
assume again that all individual costs are either zero or one. Let
$I=\{y \in V(X) \mbox{ $|$ } c_{f(y)}(y)=0 \}$ and note that $|I| \geq k$.
It can again be seen that $I$ is an independent set in $G$. \qed

We have proved the following result, conjectured in \cite{gregorykim}.

\begin{theorem}
Let $H$ be a reflexive digraph. If $H$ has a Min-Max ordering, then
MinHOM$(H)$ is polynomial time solvable; otherwise, it is NP-complete.
\end{theorem}

{\small

\end{document}